\documentclass{PoS}

\usepackage{subfig}
\usepackage{url}

\title{Parsec-scale properties of GHz-Peaked Spectrum sources from 2.3 and 8.6 GHz VLBI surveys}

\ShortTitle{Parsec-scale properties of GPS sources}

\author{Kirill V. Sokolovsky$^{1,2}$\thanks{{\bf Acknowledgments.} K.~Sokolovsky is supported by the International Max
Planck Research School (IMPRS) for Radio and Infrared Astronomy; his participation
in the $9^{th}$ EVN Symposium was partly supported by funding from the 
European Community's sixth Framework Programme under 
RadioNet R113CT 2003 5058187. Y. Y. Kovalev is a Research Fellow of the Alexander von Humboldt
Foundation. This research has made use of the NASA/IPAC Extragalactic
Database (NED)
which is operated by the Jet Propulsion Laboratory, California Institute of
Technology, under contract with the National Aeronautics and Space
Administration. The VLBA is a facility of the National Science Foundation
operated by the National Radio Astronomy Observatory under cooperative agreement
with Associated Universities, Inc. RATAN--600 observations were partly
supported by the Russian Foundation for Basic Research (projects
01-02-16812, 05-02-17377, 08-02-00545).}
~and Yuri Y. Kovalev$^{1,2}$\\
        $^{1}$Max-Planck-Institute f\"ur Radioastronomie, Auf dem H\"ugel 69, 53121 Bonn, Germany\\
        $^{2}$Astro Space Center, Lebedev Physical Institute, Profsoyuznaya 84/32, 117997 Moscow, Russia\\
        E-mail: \email{(ksokolov,ykovalev)@mpifr-bonn.mpg.de}}

\abstract{
We investigate the sample of 213 GPS sources selected from simultaneous multi-frequency
1--22~GHz observations obtained with RATAN--600 radio telescope. 
We use publicly available data to characterize parsec-scale structure 
of the selected sources. Among them we found 121 core dominated sources, 76
Compact Symmetric Object (CSO) candidates (24 of them are highly probable), 16
sources have complex parsec-scale morphology.
Most of GPS galaxies are characterized by CSO-type 
morphology and lower observed peak frequency ($\sim 1.8$~GHz). Most of GPS quasars are characterized by
``core-jet''-type morphology and higher observed peak frequency ($\sim 3.6$~GHz). This is in good agreement with previous results. 
However, we found a number of sources for which the general relation CSO -- galaxy,
core-jet -- quasar does not hold. These sources deserve detailed investigation. 
Assuming simple synchrotron model of a homogeneous cloud we estimate 
characteristic magnetic field in parsec-scale components of GPS sources to
be $B \sim 10$~mG.}

\FullConference{The 9th European VLBI Network Symposium on The role of VLBI in the Golden Age for Radio Astronomy and EVN Users Meeting\\
		 September 23-26, 2008\\
		 Bologna, Italy}

\begin{document}

\begin{figure}[!h]
\centering
\subfloat[Spectral indices of individual components]{\label{spi:spi-2}\includegraphics[height=0.34\textwidth,angle=0]{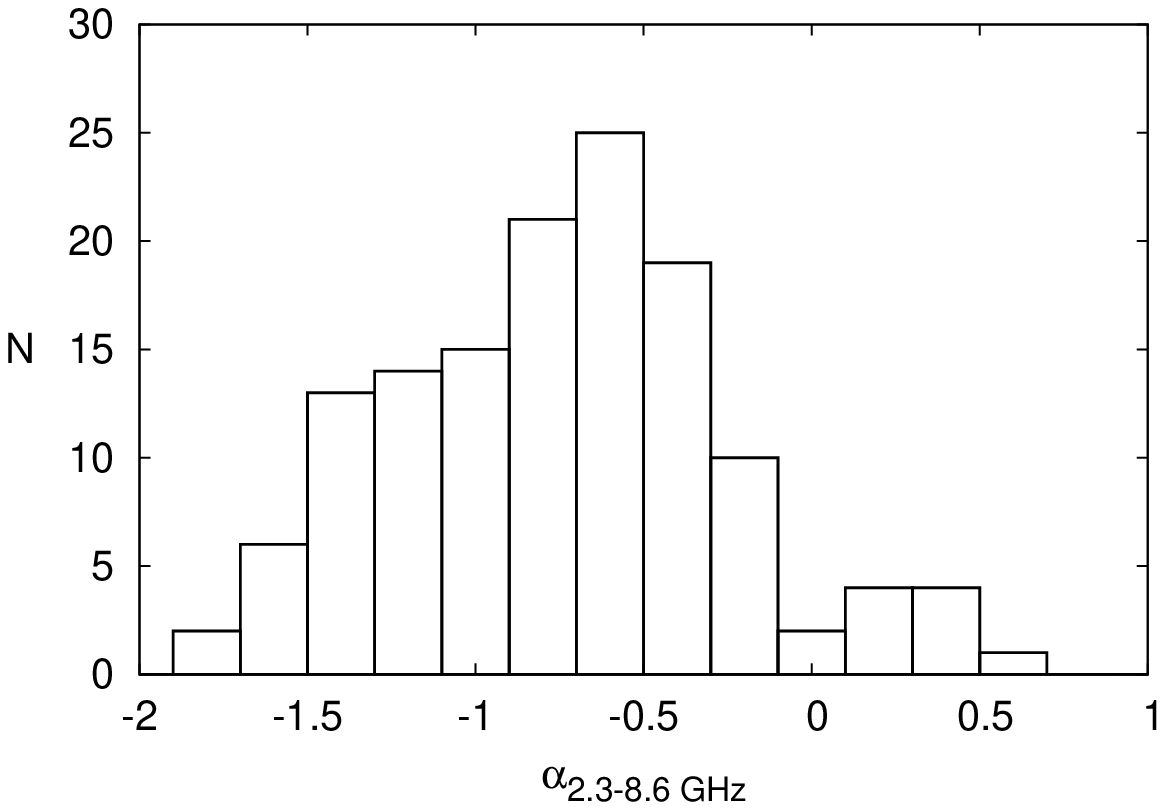}}
\subfloat[Difference between spectral indices of components]{\label{spi:spi-raznost}\includegraphics[height=0.34\textwidth,angle=0]{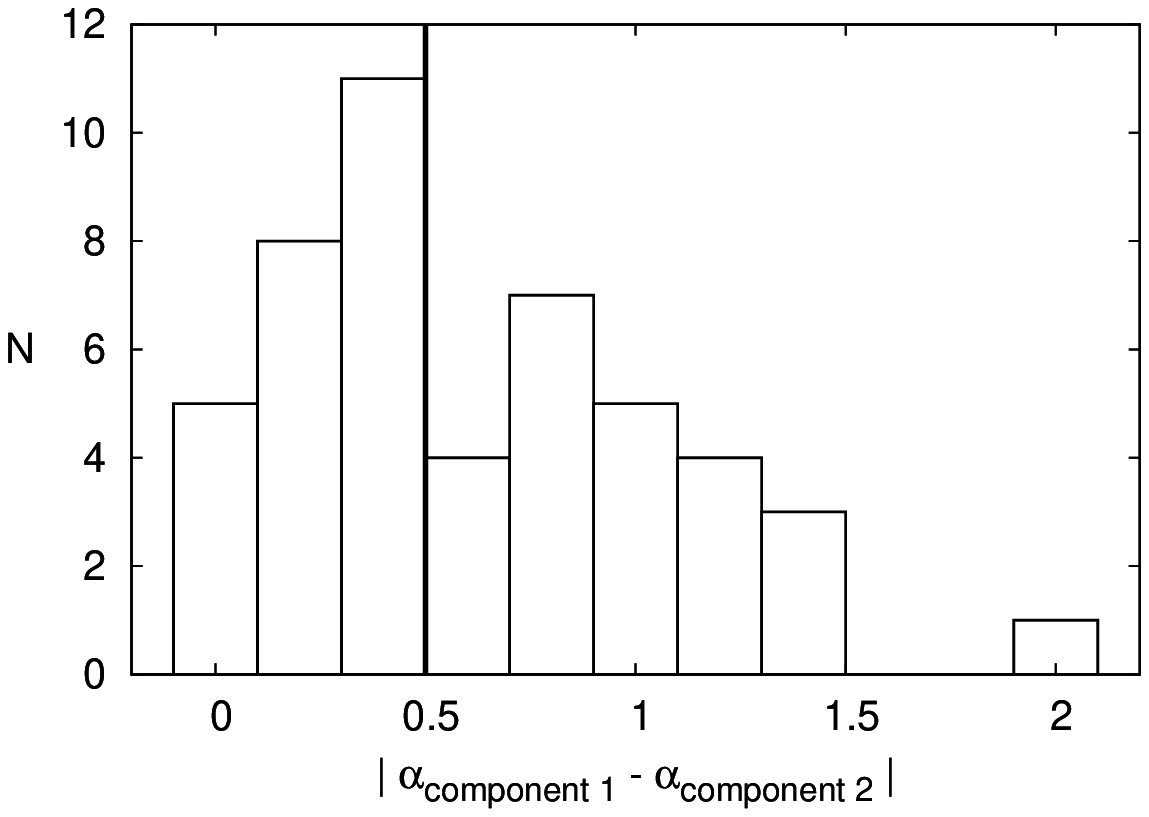}}
\caption{Distributions of 2.3--8.6~GHz spectral indices of VLBI components of CSO candidates.}
\label{spi}
\end{figure}

\begin{figure}[!h]
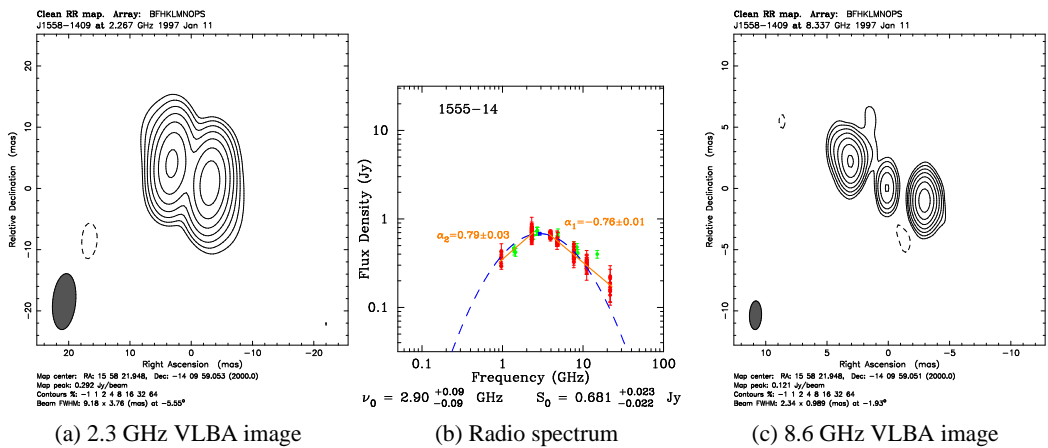

\centering
\subfloat[2.3~GHz VLBA image]{\label{fig:J1558-1409_S}\includegraphics[width=0.3\textwidth,angle=0]{J1558-1409_S.ps}}
\subfloat[Radio spectrum]{\label{fig:1555m14}\includegraphics[width=0.3\textwidth,angle=0]{1555m14.ps}}
\subfloat[8.6 GHz VLBA image]{\label{fig:J1558-1409_X}\includegraphics[width=0.3\textwidth,angle=0]{J1558-1409_X.ps}}
\caption{PKS 1555$-$140 -- new GPS galaxy with CSO morphology at $z = 0.097$. Red points on
the pannel (b) denote RATAN--600 observations, green points represent literature data.}
\label{fig}
\end{figure}

\begin{figure}[!h]
\centering
\includegraphics[height=0.34\textwidth,angle=0]{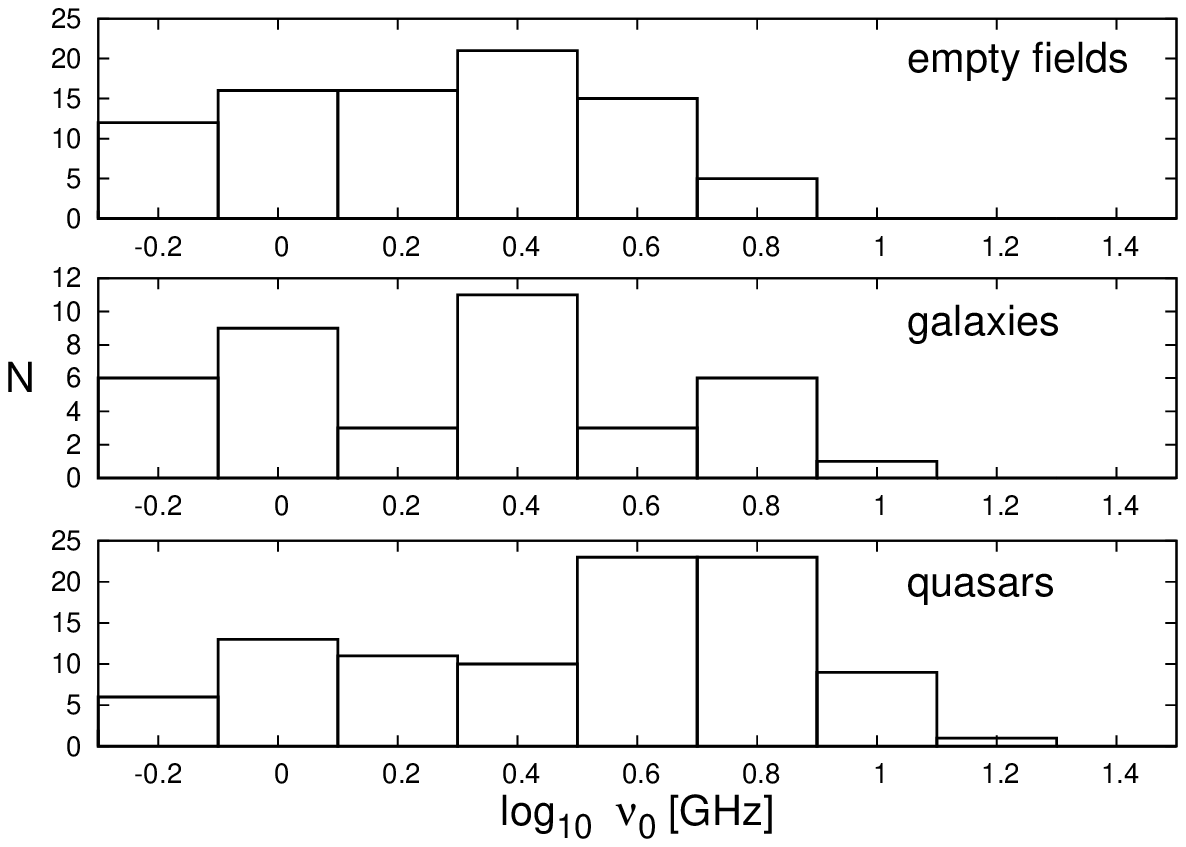}
\includegraphics[height=0.34\textwidth,angle=0]{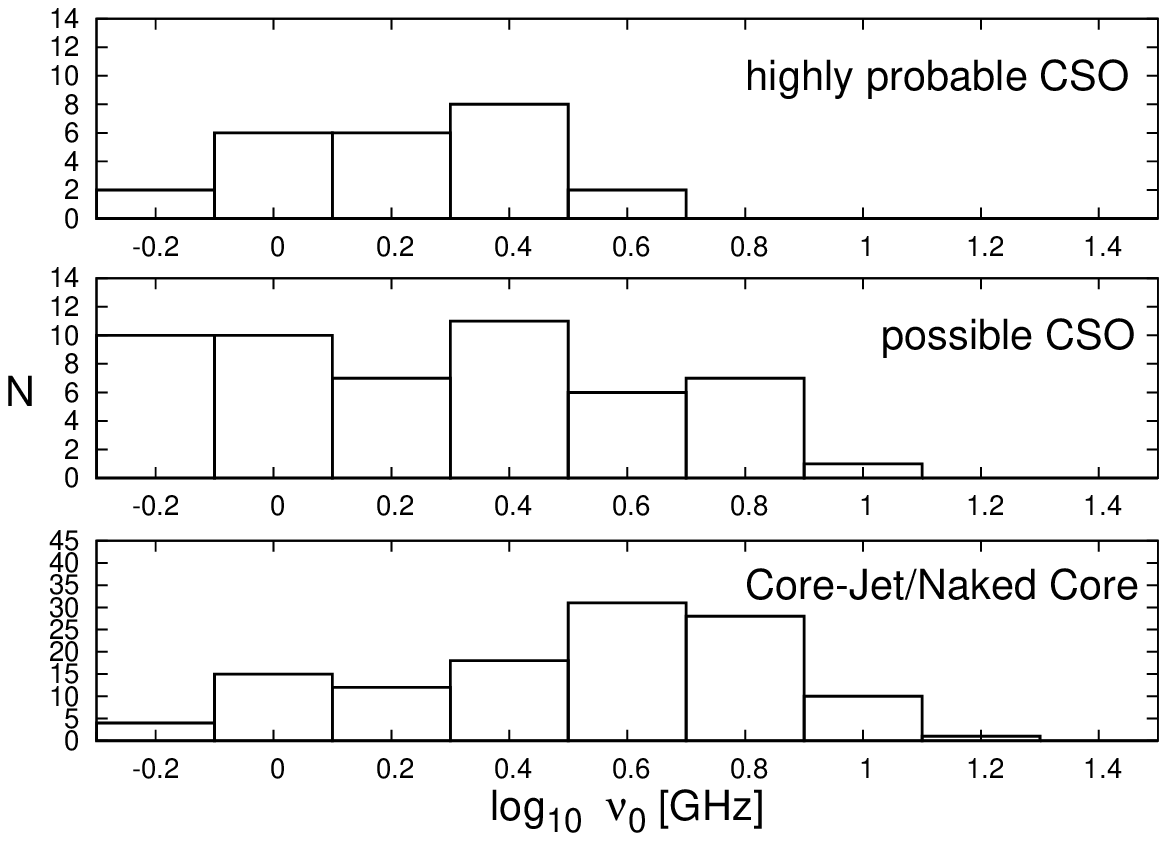}
\caption{Observed peak frequency distribution for GPS sources with different
optical identification and parsec-scale radio structure.}
\label{pfreq}
\end{figure}

\newpage

We use publicly available\footnote{Compilation of publicly available VLBI
data at \url{http://lacerta.gsfc.nasa.gov/vlbi/images}} 
2.3 and 8.6~GHz VLBI data from the VLBA Calibrator Survey
(\cite{VCS6} and references therein) and the Research and Development --
VLBA (RDV, \cite{RDV1}, \cite{Pus08}) project to characterize pc-scale properties of 213 GPS sources
from the RATAN--600 sample \cite{Sok8}. After a visual inspection of the VLBI
images we have divided the sources in to three groups: 1) ``core-jet/naked
core'' sources, 2) possible Compact Symmetric Objects (CSO) and 3) sources with
complex pc-scale morphology.

To further distinguish between true CSO and core-jet sources with two
dominating components (a core and a jet feature), we have modeled the
visibility data for all CSO candidates using {\em DIFMAP} package \cite{Difm} with
two circular Gaussian components. We
have selected sources which have two components detected at both 2.3 and
8.6~GHz images and constructed the spectral indices (between 2.3 and 8.6~GHz)
for each component (Fig.~\ref{spi:spi-2}). The distribution of difference between the spectral
indices
of two components of these sources is presented on Fig.~\ref{spi:spi-raznost}.
Since two mini-lobes of a CSO are expected to have close spectral indices, we
selected 24 sources with spectral index difference between two pc-scale
components less than 0.5 as ``highly probable CSO candidates''. Example of a newly
identified GPS galaxy associated with a CSO is presented on
Fig.~\ref{fig}.

Information about the optical identification for the sources was taken from
\cite{VV06} and from the NASA/IPAC Extragalactic Database. We found, 
with a few exceptions, that CSOs are associated with GPS galaxies and
core-jet sources are associated with GPS quasars. This is in good
agreement with previous results (e.g., \cite{Stan}). The distributions of the observed peak
frequency for GPS sources with different optical counterparts and with
different pc-scale morphology are presented on Fig.~\ref{pfreq}. GPS
sources associated with CSO are characterized by the lower observed peak
frequency ($\sim 1.8$~GHz) then core-jet sources ($\sim 3.6$~GHz).

By combining the single-dish spectrum with VLBI angular size measurements we
can estimate the magnetic field in pc-scale components of a radio source using 
simple synchrotron model (e.g., \cite{Slys}, \cite{Mar3}). We estimate the characteristic
magnetic field in the pc-scale components of GPS sources to be $B \sim 10$~mG.

\end{document}